\documentclass[10pt,letterpaper]{article}

\usepackage{cogsci}

\cogscifinalcopy 
\usepackage{graphicx}

\usepackage{pslatex}
\usepackage{apacite}
\usepackage{float} 

\title{Semantic distance organizes social knowledge: Insights from semantic dementia and cross-modal conceptual space}

\author{Y. Ivette Col\'on, Matthew Rouse, Matthew A. Lambon Ralph, and Timothy T. Rogers}

\begin{document}

\maketitle

\begin{abstract}

Our interaction with others largely hinges on how we semantically organize the social world. The organization of such conceptual information is not static-- as we age, our experiences and ever-changing anatomy alter how we represent and arrange semantic information. How does semantic distance between concepts affect this organization, particularly for those with pathological deficits in semantic knowledge? Using triplet judgment responses collected from healthy participants, we compute an ordinal similarity embedding for a set of social words and images that vary in the dimensions of age and gender. We compare semantic distances between items in the space to patterns of error in a word-picture matching task performed by patients with semantic dementia (SD). Error patterns reveal that SD patients retain gender information more robustly than age information, and that age-related errors are a function of linear distance in age from a concept word. The distances between probed and exemplar items in the resulting conceptual map reflect error patterns in SD patient responses such that items semantically closer to a probed concept-- in gender category or in linear age-- are more likely to be erroneously chosen by patients in a word-picture matching task. To our knowledge, this is the first triplet embedding work to embed representations of words and images in a unified space, and to use this space to explain patterns of behavior in patients with impaired social semantic cognition.

\end{abstract}

\section{Introduction}

Patients with pathology in the anterior temporal lobes (ATLs) often show significant impairment to social knowledge, including meanings of social words \cite{zahn2007social}, an inability to recognize faces \cite{damasio1990face}, and disordered recognition of emotional expressions \cite{lindquist2014emotion}. It remains unclear what such patterns suggest about the organization of neural systems that support social cognition. One view is that the deficit arises because the ATLs constitute an important part of a cortical network dedicated specifically to social cognition. An alternative view suggests instead that the ATLs play a critical role for all manner conceptual knowledge including people, places, animals, objects, emotions, and abstract concepts \cite{ralph2017neural,balgova2022role}, so that disordered social knowledge is just one facet of a broader domain-general semantic impairment. 
The views have been difficult to adjudicate for two reasons. First, social vs semantic studies often use quite different materials, methods, and procedures, making direct comparison across existing studies difficult. Second, the behaviors of patients suffering from ATL pathology, and the patterns of ATL functional activation elicited in healthy participants, are both strongly influenced by a variety of potentially confounding factors such as concept frequency, familiarity, and concreteness \cite{rogers2015disorders, rice2018concrete}. 

One factor known to influence impaired behavior on semantic tasks is the {\it precision} with which a stimulus must be identified to support successful performance. For object and animal concepts, tasks requiring more precise identification are more vulnerable to semantic impairment arising from ATL pathology, and elicit greater functional activation in ATL regions. For instance, when matching a word such as ``robin'' to a corresponding picture, patients with semantic dementia (SD)---a progressive degenerative illness characterized by cortical thinning in the ATLs bilaterally---can perform quite well when distracting images are all semantically distal to the target (e.g. all non-animals) but show much larger impairment when distractors are semantically proximal (e.g. other small birds) \cite{rogers2007object}.

\begin{figure*}[ht!]
  \centering
    \includegraphics[width=\textwidth]{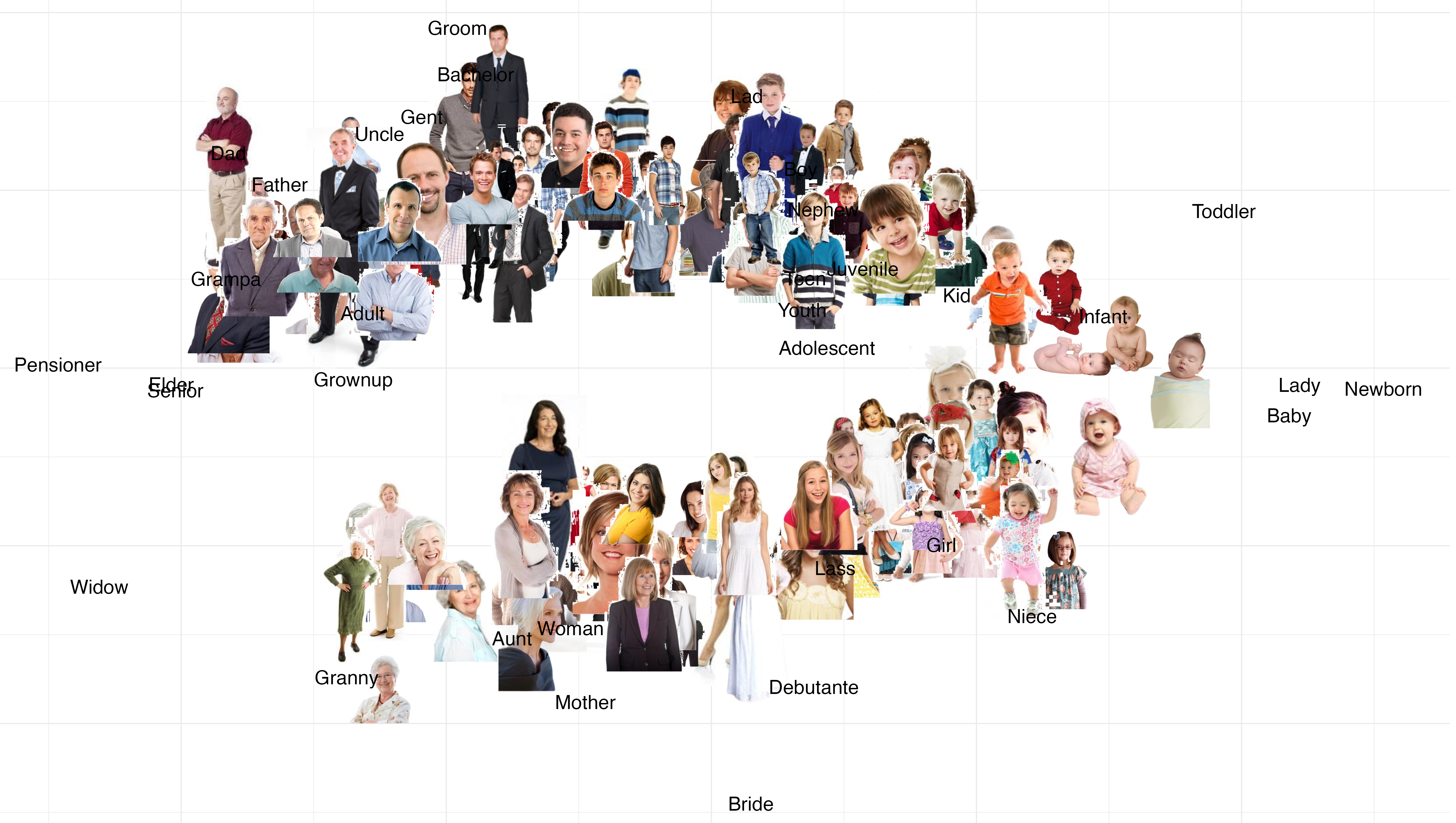}
    \caption{A two-dimensional embedding of social words and images. The horizontal dimension clearly expresses perceived age, while the vertical dimension captures perceived gender. Word embedding coordinates reside near images depicting examples of the name category.}
  \label{fig:embed}
\end{figure*}

Thus to assess whether social concepts are disproportionately dependent on ATLs relative to other non-social concepts, one requires tasks and materials in both domains that permit fine control of the precision with which a conceptual representation must be resolved for good performance. The current paper presents an initial step in this direction by first (a) measuring the conceptual similarities existing amongst social words and images and then (b) assessing whether and how such structure influences the ability of patients with ATL pathology to match words to corresponding pictures amidst varying kinds of distractors. The results suggest that social concepts, like object concepts, can be characterized as residing within a continuous representational space, and that the social knowledge impairments observed with ATL damage are strongly governed by such structure. 

\section{Study 1: Measuring conceptual structure in social stimuli}

The goal of Study 1 was to establish a semantic representational space for a set of social agent stimuli, including both words and images, such that the distances between pairs of items in the space is proportional to their perceived semantic relatedness. Such a space can allow us to empirically measure the semantic distance between words and images, providing a basis for assessing whether such distances influence patients' ability to match words to semantically corresponding pictures. 

By {\it social agent stimuli}, we mean words denoting various categories of people (e.g. "grampa", "toddler," "widow," etc.) and photographs of people belonging to these categories. Many such words denote a person's perceived gender and/or age, so the images include pictures of people across the life span from infants to senior citizens and with equal proportions of male- vs female-appearing individuals. 

For object concepts, semantic representational spaces are commonly estimated by asking participants to list properties of various items and measuring the degree of feature overlap between items \cite{mcrae2005semantic,de2008exemplar}. It is less clear that such an approach will extend well to social concepts, so we pursued an alternative approach, namely triadic comparisons \cite{jamieson2011low}. On each trial of the task the participant views a {\it sample} stimulus and two {\it choice} stimuli, and must decide which of the choice stimuli is most similar to the sample. From many such judgments, the stimuli can be embedded in a low-dimensional space so that pairs of items often chosen together as ``more similar'' relative to some third item are situated nearby. The semantic distance between any two items is then measured as the Euclidean distance between coordinates of the corresponding points in the embedding space. Thus triadic comparisons offer a means of estimating semantic distances amongst both words and images, 
for comparatively abstract social concepts such as "adult" or "debutante", and without requiring participants to list features.

In this study, we embedded both words and images together within a common semantic space using the triadic comparison task. This joint, cross-modal embedding provides a basis for measuring semantic distances between a word representation and various image representations, providing a quantitative basis for assessing how precisely a participant must resolve a word-meaning representation in order to correctly choose a matching image from an array of distractors. That is, the joint embedding should provide a basis for predicting patterns of errors on word-picture matching tasks similar to those commonly used to study object semantics. To our knowledge, this is the first triplet judgment study to embed images \textit{and} words in a shared conceptual space. 

\subsection{Methods}

    \textbf{Stimuli.} 
    The stimuli include words and pictures taken from a word-picture matching task used in a broader study of social concepts in semantic dementia \cite{rouse2023neuropsychological}. They include 35 target words, all of which denote categories of people and varying in their implied gender and/or age categories (e.g., "lad", "bride", "elder," etc), as well as 140 images, including the target and three distractor images for each of the 35 words. The same 35 concept words and 140 images used in the triplet judgment task were also used in the social word-picture matching task. 

\textbf{Behavioral Procedure.} We ran an online triadic comparison task with 510 trials per participant (a block of 255 social trials including 5 attention-check trials, and a block of 255 non-social trials including 5 check trials). Participants completed either a block of 255 social item trials or 255 nonsocial item trials, followed by a short break, the remaining block (whichever domain they had not previously seen), and a brief demographic questionnaire. The present work focuses only on the social block of this task. 

In each trial of the triadic comparison task participants saw three items: a reference item, and two choice items beneath the reference item. Respondents were instructed to  "select the item most like the item on top" for each trial by clicking on one of the two choice items. The majority of trials (200 out of 255) were randomly-generated combinations of images and words. Each participant also viewed a set of 50 {\it validation} triplets randomly sampled from a pool of 200, ensuring that some triplets were viewed by multiple participants and thus allowing us to compute inter-rater agreement from these trials. To ensure attention during the task, participants also viewed 5 attention-check trials interspersed throughout the block, where one of the choice items was identical to the target items. If participants made an incorrect decision on check trials, or if their mean response times dropped below 1s, they received feedback asking them to pay closer attention.

\begin{figure*}[ht!]
  \centering
    \includegraphics[scale=0.6, width=\textwidth]{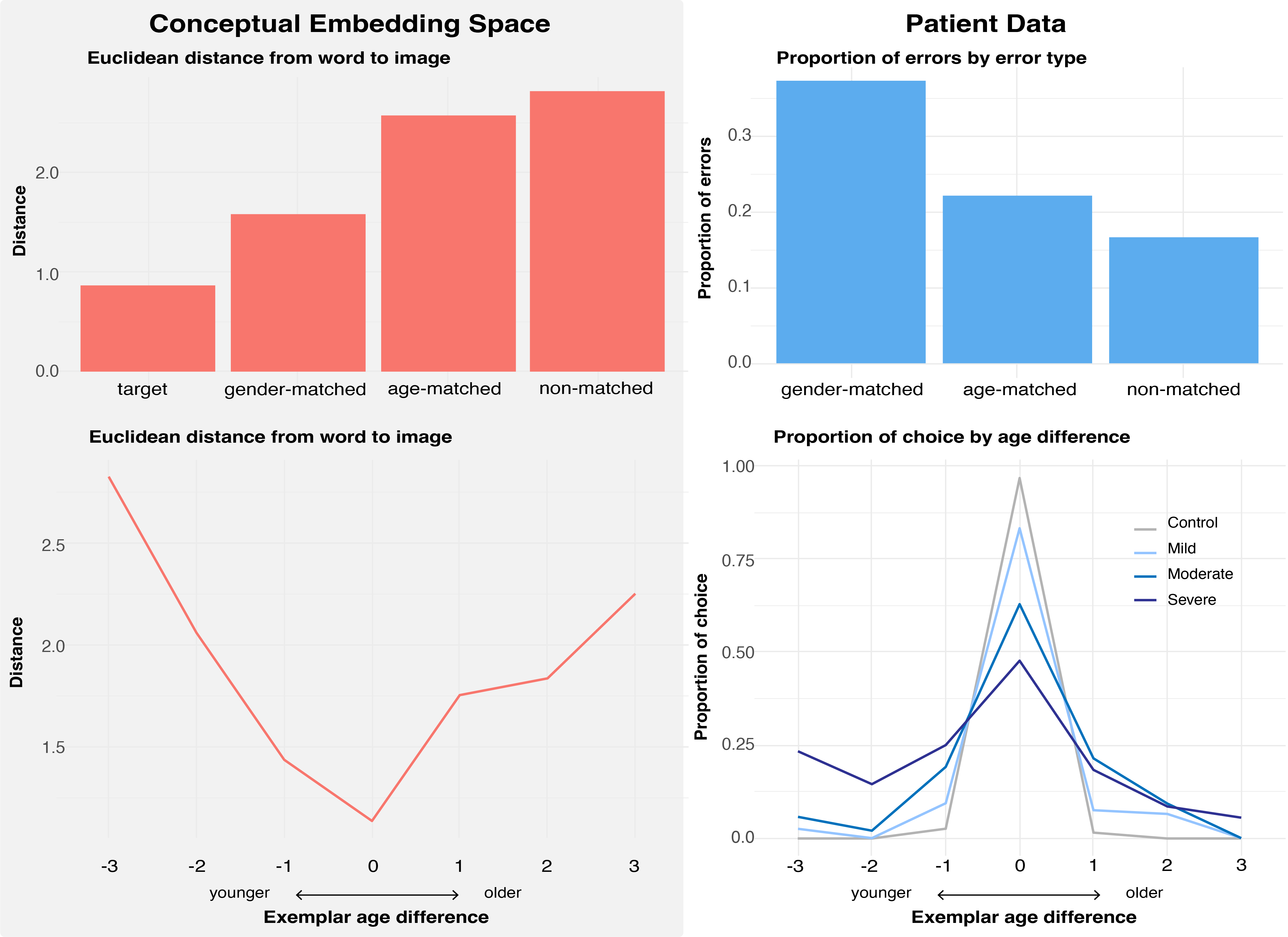}
    \caption{Comparison between patients and conceptual embedding distances. Gender-matched test items are closest to the target item in embedding distance and account for the highest proportion of errors in the patient data. Distance in the embedding space increases as distance in age from a target item increases, reflecting the distribution of patient errors for linear age trials.}
  \label{fig:compare}
\end{figure*}

\textbf{Population.} Previous work throughout the fields of psychology and anthropology has established that conceptual organization can vary with the cultural experiences of a given population. As we are interested in a conceptual space that could reasonably explain patterns of error from semantic dementia patients residing in the UK, we wanted to compute a conceptual space from participants as closely matched in age and culture to the studied patients as possible. To that end, we recruited 50 older adults from Prolific who were living in the UK and were older than 50 years of age (Range = 50-79, M = 61 years; 66\% female). Of the recruited participants, all participants passed our performance-based exclusion criteria (average RT $>$ 1 second per trial and at least $\geq$\ 80\% correct on check trials). We removed one participant due to incomplete data, resulting in 49 participants for analysis.

\textbf{Computing embeddings.} We compiled all triadic comparison judgments for social items and computed a two-dimensional embedding using the OfflineEmbedding function in the Salmon Python package \cite{sievert2023efficiently}. As with other ordinal embedding algorithms, this function generates an n-dimensional similarity embedding from triadic judgments through a sampled training and testing model. For this model, we used prediction error on a held-out test set of triplets (20\% of responses) to choose the best embedding across epochs, selecting a final embedding when prediction error did not improve for 10,000 epochs across the maximum 30,000 training epochs. Since our stimuli vary primarily in age and perceived/implied gender, we embedded the stimuli into two dimensions. For embeddings in higher dimensions (3, 4, and 5), a principle component analysis found that 98 percent of variance in pairwise distances were explained by just two components, suggesting that the additional dimensionality was not encoding important structure.

\subsection{Results}

Figure \ref{fig:embed} shows the 2D embedding for the words and images recovered from the triadic comparisons. Three features of the plot are worth noting. First, the space is strongly shaped by age and gender. Second, by inspection, coordinates for most words lie quite close to corresponding images. Third, the distance between perceived genders is quite wide for older adults and much narrower for young children--perhaps indicating the relatively lesser importance of gender in representations of young children in this older age cohort. Fourth, words that are ostensibly gender-neutral (e.g. ``adult'', ``youth'', ``kid'') reside within the ``male'' part of the representation space; only words explicitly denoting femaleness align with images of female-appearing people. Thus, within this cohort, gender-neutral terms may be understood as indicating maleness, reflecting recent postulations of androcentrism as a social cognitive default \cite{bailey2019man}. 

The 2D embedding allows us to compute, for each item in the social word-picture matching task, the semantic distance between the test word and the four choice images, including the target and the three distractors. Clearly the target image should be closer to the test word than the three distractors; however the embedding allows us to consider which distractors should be most difficult to discriminate from the target.

\textbf{Gender-matched items closer to concept than age-matched items.} For test words that imply both age and gender (e.g. ``grampa''), the distractors include an age-matched item (e.g. an older woman), a gender-matched item (e.g. a younger man), and a non-match (e.g. a younger woman). For these items, distances in the computed two-dimensional embedding reveal a consistent pattern: for any given concept word, corresponding gender-matched images are closer than the concept's age-matched images, and both are closer still than images that are unmatched in both age and gender (see Figure \ref{fig:compare}, top). In other words, the space suggests that for a given concept, the dimension of gender is more likely to be preserved under impairment than the dimension of age, leading to potential errors in choosing gender-matched distractors over age-matched distractors. 

\textbf{Greater difference in age leads to greater representational distance.} For test words that imply age but not gender (e.g. ``infant''), distractor images always matched the target image in gender, but varied in how distant they were from the target age. To measure this distance, we assigned each image to one of 6 age brackets: infant, toddler, child, teen, adult, or senior. For each test array, we then counted how many brackets away from the target image each distractor was. For instance, if the target image was a teen, a distractor image of a child was coded as -1 (one bracket younger), while a distractor image of a senior was coded as +2 (two brackets older). Using this coding scheme, we computed the mean semantic representational distance of distractors within each level of age difference.

The results are shown in Figure \ref{fig:compare} (bottom). As the distactor image grows increasingly distal to the target in age, it also grows increasingly distal in the embedding space. Thus if social knowledge impairments are governed by this representational structure, patients should more often choose distractor images that are close to the target image in perceived age.  

\textbf{Discussion.} Study 1 shows that words and images denoting social concepts can be embedded within a common representational semantic space using triadic comparison judgments. When judgments are collected from a healthy older UK population, the resulting embedding reveals a remarkable alignment between word meanings and corresponding images. The structure of this space implies that this population organizes social concepts largely with respect to age and perceived gender; that gender is more highly salient for older people; and that gender-neutral social words are represented as indicating male-appearing people. The study further shows how embeddings support computation of representational distances from a word to various pictures, leading to specific predictions about patterns of impairment in the social word-picture matching task. Specifically, for words implying both age and gender, patients with social conceptual impairments should make more errors to distractors that are matched for gender and different in age than vice versa, because such images are closer in the representational space. For gender-neutral words, participants should make more errors to distractors that are close in age to the target image. These are the predictions tested in Study 2.

\section{Study 2: Social word-picture matching in semantic dementia}

The data for Study 2 were collected as part of a much broader study of social conceptual knowledge in patients with fronto-temporal dementia \cite{rouse2023neuropsychological}. In that study, 21 patients diagnosed with semantic dementia and 19 age-matched controls were tested on a large battery of cognitive tasks which included a range of standard clinical assessments as well as tests designed to test both social and non-social conceptual knowledge. Patients were diagnosed according to consensus criteria including impaired naming, impaired verbal and nonverbal comprehension, other aspects of cognition (perception, memory, language fluency, executive function) normal or near normal at first assessment, and MRI-confirmed pathology and/or hypometabolism in the anterior temporal cortex. The broader study collected many measures from each patient over a period of several months; here we focus only on the social word-picture matching task that provided the stimuli used in Study 1.

 \subsection{Methods}
 
    \textbf{Participants.} 
    Of the  21 patients with semantic dementia included in this analysis, eight patients were diagnosed with a mild progression of the disease, seven with moderate disease, and six with severe disease. In addition, 19 healthy age-matched adults from the same geographic region completed the task as controls. All patients lived in the UK and were between ages of 50-85.   

    \textbf{Stimuli.}
    Stimuli were the same items used in Study 1 from the social word-picture matching task that includes 35 social concept words, each appearing with a target image and three distractor images (140 images total).
    Target and distractor images correspond to gender or age categories relevant to the probed concept word. Some images depict additional socially-relevant information through their clothing (e.g., the "bride" target image depicts a woman in a white dress).

    \textbf{Social word-picture matching}. 
     We focus on patterns of error in the social word-picture matching task. In this task, there are two distinct trial types capturing different organizational principles. The first are \textit{orthogonal-match trials}, where we manipulate the similarity of items by relatively orthogonal dimensions of (binarized) gender and age. In orthogonal-match trials, participants are presented with a concept word (for example, "Bride"), and four possible choice images. One image is intended as the target exemplar for the concept (in this case, an adult woman), and the other three are controlled foils: one image is matched in gender to the target concept but differs in age (e.g., an adolescent girl); one image is matched in age but differs in gender (e.g., an adult man); and one image that shares neither dimension (e.g., a adolescent boy). Note that, although in reality gender is a continuous dimension, we use the binary categories of man/boy and woman/girl in this task.

     By contrast, \textit{linear-age trials} allow us to manipulate item similarity along a single dimension. In linear-age trials, participants are again shown a concept word and four image choices, but here the distractor images are in different relative age brackets from the target image. For example, if the concept word is "toddler", the target image of a toddler would be at 0 on this relative continuum. An age bracket younger (infants) would be -1 step; two age brackets older (teen) would be +2 steps. Our range of age categories includes infants, toddlers, adolescents, teens, adults, and seniors. Note that physical and intellectual changes occur more rapidly in early life, leading to more parcelized conceptual age brackets for younger people. 

    \textbf{Procedure.} Patients were tested in a quiet room. The test began with three easy practice trials to ensure the participant understood the task. On each trial the experimenter showed the participant a display with the test word printed at the top and four images below. The experimenter read the word aloud and asked the patient to point to the image that best matched the word. The experimenter recorded which item was selected on each trial without providing feedback to the participant. The location of the target image was counterbalanced across items. Test items were presented in the same randomly-determined order to all participants.

\subsection{Results}

    Healthy controls performed at or near ceiling on the task, scoring 98\% correct on average. All patients were significantly impaired on the task, with performance for each falling more than two standard deviations below the control mean. The two best-performing patients, both diagnosed with mild disease, performed as well as the worst-performing control, scoring 92\% correct. The magnitude of the impairment was larger in patients with more severe disease (mean of 90\% correct for mild patients, 82\% for moderate patients, and 59\% for severe patients).

    The key interest for the current work concerns patterns of errors committed in the different trial types. For orthogonal-match trials, we computed the proportion of errors committed to each distractor type: gender-match, age-match, or non-match. For linear-age trials, we computed the proportion of errors committed to each level of age-difference among the distractors. These data are shown in Figure \ref{fig:compare}. 
    
    The top plot shows a mirror-image of the results from Study 1: more errors are committed to distractors that are closer together in the semantic representation space. Thus the patients more often choose the gender-matched item than the age-matched and non-match item--effectively preserving the more robust gender dimension. Likewise for linear-age trials, across levels of severity participants committed more errors toward distractors nearby in age. As disease severity increased, however, participants grew increasingly likely to erroneously choose distractors more distal to the target in age---as though their mental representation of the age implied by the word is growing increasingly blurry.

\section{General Discussion}

In this work we measured conceptual similarity amongst a set of social words and images by situating them in a cross-modal social semantic space. We then used the resulting space to evaluate patterns of social-knowledge impairment in patients with ATL pathology. Consistent
with prior work on object concepts, the results indicate that performance on a social word-picture matching task is profoundly influenced by the semantic similarity relations existing amongst the test word, the target image, and the distracting images. Items proximal to a probed concept, be it in gender category or linear age, were more prone to attract errors. Notably, gender-matched items consistently exhibited closer proximity to concept words than age-matched items in the embeddings from Study 1, and accounted for a higher proportion of errors in patient word-picture matching in Study 2. Thus the relative preservation of gender over age information in the disorder may reflect the greater salience/importance of this dimension for organizing social concepts in this age and culture cohort.

The disparity in organization between gender and age dimensions potentially reflects differences in representational flexibility inherent in these dimensions. While aging manifests universally and implies change, gender is often perceived as a relatively stable characteristic that is less prone to change over time, a fact which may be attributed to the historical conception of gender as a binary construct in contrast to the multifaceted nature of human development across various life stages. These findings underscore the necessity of precise representations for successful social categorization, where dimensions culturally and conceptually perceived as distinct, such as gender, are likely represented more robustly within a concept.  It is worth noting, however, that measurements of conceptual organization are highly dependent on the particular sample of respondents, such that a sample differing in age or culture (say, American undergraduates) may reflect different organization of attributes like gender in their responses and may not reflect the androcentrism suggested by the current sample.

Although our study primarily focused on conceptual knowledge of social agents, the dimension of age extends beyond the social realm to encompass all entities, including people, objects, and ideas. The differential patterns observed between age and gender in our data challenge a view of ATL function as exclusively dedicated to social conceptual knowledge. Instead they suggest that, while ATL pathology may appear to impact some kinds of information over others, such phenomena may simply reflect the structure and organization of the healthy representational space as it begins to degrade. Future investigations into non-social semantic organization in SD patients will further clarify the principles underlying category formation across domains. 

The results are consistent with the view that the ATLs form a domain-general and trans-modal semantic hub responsible for encoding conceptual similarity structure and connecting various sensory, motor, linguistic, and affective representations distributed in cortex. This study successfully integrates verbal and pictorial representations within a unified semantic space, allowing for computation of semantic distances amongst social stimuli presented across language and visual-object modalities. The multi-modal conceptual space in this work is a step towards aligning semantic spaces generated from behavior to the complex functioning of the region we consider responsible for representing semantic knowledge. 

We anticipate exploring the generalizability of these findings to higher-dimensional semantic contexts encompassing both social and non-social dimensions. Additionally, comparative analysis of error patterns across various semantic domains may elucidate domain-specific mechanisms underlying semantic deterioration in neurodegenerative diseases, though we make no claims that semantic cognition is necessarily carved into functionally distinct, domain-specific processes. Still, our findings offer insights into the organization and degradation of social conceptual knowledge in semantic dementia, and emphasizes the importance of semantic distance and cross-modal representations in understanding cognitive decline.

\bibliographystyle{apacite}

\setlength{\bibleftmargin}{.125in}
\setlength{\bibindent}{-\bibleftmargin}

\bibliography{CogSci_Template}

\begin{thebibliography}{}

\bibitem [\protect \citeauthoryear {%
Bailey%
, LaFrance%
\BCBL {}\ \BBA {} Dovidio%
}{%
Bailey%
\ \protect \BOthers {.}}{%
{\protect \APACyear {2019}}%
}]{%
bailey2019man}
\APACinsertmetastar {%
bailey2019man}%
\begin{APACrefauthors}%
Bailey, A\BPBI H.%
, LaFrance, M.%
\BCBL {}\ \BBA {} Dovidio, J\BPBI F.%
\end{APACrefauthors}%
\unskip\
\newblock
\APACrefYearMonthDay{2019}{}{}.
\newblock
{\BBOQ}\APACrefatitle {Is man the measure of all things? A social cognitive account of androcentrism} {Is man the measure of all things? a social cognitive account of androcentrism}.{\BBCQ}
\newblock
\APACjournalVolNumPages{Personality and social psychology review}{23}{4}{307--331}.
\PrintBackRefs{\CurrentBib}

\bibitem [\protect \citeauthoryear {%
Balgova%
, Diveica%
, Walbrin%
\BCBL {}\ \BBA {} Binney%
}{%
Balgova%
\ \protect \BOthers {.}}{%
{\protect \APACyear {2022}}%
}]{%
balgova2022role}
\APACinsertmetastar {%
balgova2022role}%
\begin{APACrefauthors}%
Balgova, E.%
, Diveica, V.%
, Walbrin, J.%
\BCBL {}\ \BBA {} Binney, R\BPBI J.%
\end{APACrefauthors}%
\unskip\
\newblock
\APACrefYearMonthDay{2022}{}{}.
\newblock
{\BBOQ}\APACrefatitle {The role of the ventrolateral anterior temporal lobes in social cognition} {The role of the ventrolateral anterior temporal lobes in social cognition}.{\BBCQ}
\newblock
\APACjournalVolNumPages{Human Brain Mapping}{43}{15}{4589--4608}.
\PrintBackRefs{\CurrentBib}

\bibitem [\protect \citeauthoryear {%
Damasio%
, Tranel%
\BCBL {}\ \BBA {} Damasio%
}{%
Damasio%
\ \protect \BOthers {.}}{%
{\protect \APACyear {1990}}%
}]{%
damasio1990face}
\APACinsertmetastar {%
damasio1990face}%
\begin{APACrefauthors}%
Damasio, A\BPBI R.%
, Tranel, D.%
\BCBL {}\ \BBA {} Damasio, H.%
\end{APACrefauthors}%
\unskip\
\newblock
\APACrefYearMonthDay{1990}{}{}.
\newblock
{\BBOQ}\APACrefatitle {Face agnosia and the neural substrates of memory} {Face agnosia and the neural substrates of memory}.{\BBCQ}
\newblock
\APACjournalVolNumPages{Annual review of neuroscience}{13}{1}{89--109}.
\PrintBackRefs{\CurrentBib}

\bibitem [\protect \citeauthoryear {%
De~Deyne%
\ \protect \BOthers {.}}{%
De~Deyne%
\ \protect \BOthers {.}}{%
{\protect \APACyear {2008}}%
}]{%
de2008exemplar}
\APACinsertmetastar {%
de2008exemplar}%
\begin{APACrefauthors}%
De~Deyne, S.%
, Verheyen, S.%
, Ameel, E.%
, Vanpaemel, W.%
, Dry, M\BPBI J.%
, Voorspoels, W.%
\BCBL {}\ \BBA {} Storms, G.%
\end{APACrefauthors}%
\unskip\
\newblock
\APACrefYearMonthDay{2008}{}{}.
\newblock
{\BBOQ}\APACrefatitle {Exemplar by feature applicability matrices and other Dutch normative data for semantic concepts} {Exemplar by feature applicability matrices and other dutch normative data for semantic concepts}.{\BBCQ}
\newblock
\APACjournalVolNumPages{Behavior research methods}{40}{}{1030--1048}.
\PrintBackRefs{\CurrentBib}

\bibitem [\protect \citeauthoryear {%
Jamieson%
\ \BBA {} Nowak%
}{%
Jamieson%
\ \BBA {} Nowak%
}{%
{\protect \APACyear {2011}}%
}]{%
jamieson2011low}
\APACinsertmetastar {%
jamieson2011low}%
\begin{APACrefauthors}%
Jamieson, K\BPBI G.%
\BCBT {}\ \BBA {} Nowak, R\BPBI D.%
\end{APACrefauthors}%
\unskip\
\newblock
\APACrefYearMonthDay{2011}{}{}.
\newblock
{\BBOQ}\APACrefatitle {Low-dimensional embedding using adaptively selected ordinal data} {Low-dimensional embedding using adaptively selected ordinal data}.{\BBCQ}
\newblock
\BIn{} \APACrefbtitle {2011 49th Annual Allerton Conference on Communication, Control, and Computing (Allerton)} {2011 49th annual allerton conference on communication, control, and computing (allerton)}\ (\BPGS\ 1077--1084).
\PrintBackRefs{\CurrentBib}

\bibitem [\protect \citeauthoryear {%
Lambon~Ralph%
, Jefferies%
, Patterson%
\BCBL {}\ \BBA {} Rogers%
}{%
Lambon~Ralph%
\ \protect \BOthers {.}}{%
{\protect \APACyear {2017}}%
}]{%
ralph2017neural}
\APACinsertmetastar {%
ralph2017neural}%
\begin{APACrefauthors}%
Lambon~Ralph, M\BPBI A.%
, Jefferies, E.%
, Patterson, K.%
\BCBL {}\ \BBA {} Rogers, T\BPBI T.%
\end{APACrefauthors}%
\unskip\
\newblock
\APACrefYearMonthDay{2017}{}{}.
\newblock
{\BBOQ}\APACrefatitle {The neural and computational bases of semantic cognition} {The neural and computational bases of semantic cognition}.{\BBCQ}
\newblock
\APACjournalVolNumPages{Nature reviews neuroscience}{18}{1}{42--55}.
\PrintBackRefs{\CurrentBib}

\bibitem [\protect \citeauthoryear {%
Lindquist%
, Gendron%
, Barrett%
\BCBL {}\ \BBA {} Dickerson%
}{%
Lindquist%
\ \protect \BOthers {.}}{%
{\protect \APACyear {2014}}%
}]{%
lindquist2014emotion}
\APACinsertmetastar {%
lindquist2014emotion}%
\begin{APACrefauthors}%
Lindquist, K\BPBI A.%
, Gendron, M.%
, Barrett, L\BPBI F.%
\BCBL {}\ \BBA {} Dickerson, B\BPBI C.%
\end{APACrefauthors}%
\unskip\
\newblock
\APACrefYearMonthDay{2014}{}{}.
\newblock
{\BBOQ}\APACrefatitle {Emotion perception, but not affect perception, is impaired with semantic memory loss.} {Emotion perception, but not affect perception, is impaired with semantic memory loss.}{\BBCQ}
\newblock
\APACjournalVolNumPages{Emotion}{14}{2}{375}.
\PrintBackRefs{\CurrentBib}

\bibitem [\protect \citeauthoryear {%
McRae%
, Cree%
, Seidenberg%
\BCBL {}\ \BBA {} McNorgan%
}{%
McRae%
\ \protect \BOthers {.}}{%
{\protect \APACyear {2005}}%
}]{%
mcrae2005semantic}
\APACinsertmetastar {%
mcrae2005semantic}%
\begin{APACrefauthors}%
McRae, K.%
, Cree, G\BPBI S.%
, Seidenberg, M\BPBI S.%
\BCBL {}\ \BBA {} McNorgan, C.%
\end{APACrefauthors}%
\unskip\
\newblock
\APACrefYearMonthDay{2005}{}{}.
\newblock
{\BBOQ}\APACrefatitle {Semantic feature production norms for a large set of living and nonliving things} {Semantic feature production norms for a large set of living and nonliving things}.{\BBCQ}
\newblock
\APACjournalVolNumPages{Behavior research methods}{37}{4}{547--559}.
\PrintBackRefs{\CurrentBib}

\bibitem [\protect \citeauthoryear {%
Rice%
, Hoffman%
, Binney%
\BCBL {}\ \BBA {} Lambon~Ralph%
}{%
Rice%
\ \protect \BOthers {.}}{%
{\protect \APACyear {2018}}%
}]{%
rice2018concrete}
\APACinsertmetastar {%
rice2018concrete}%
\begin{APACrefauthors}%
Rice, G\BPBI E.%
, Hoffman, P.%
, Binney, R\BPBI J.%
\BCBL {}\ \BBA {} Lambon~Ralph, M\BPBI A.%
\end{APACrefauthors}%
\unskip\
\newblock
\APACrefYearMonthDay{2018}{}{}.
\newblock
{\BBOQ}\APACrefatitle {Concrete versus abstract forms of social concept: an fMRI comparison of knowledge about people versus social terms} {Concrete versus abstract forms of social concept: an fmri comparison of knowledge about people versus social terms}.{\BBCQ}
\newblock
\APACjournalVolNumPages{Philosophical Transactions of the Royal Society B: Biological Sciences}{373}{1752}{20170136}.
\PrintBackRefs{\CurrentBib}

\bibitem [\protect \citeauthoryear {%
Rogers%
\ \BBA {} Patterson%
}{%
Rogers%
\ \BBA {} Patterson%
}{%
{\protect \APACyear {2007}}%
}]{%
rogers2007object}
\APACinsertmetastar {%
rogers2007object}%
\begin{APACrefauthors}%
Rogers, T\BPBI T.%
\BCBT {}\ \BBA {} Patterson, K.%
\end{APACrefauthors}%
\unskip\
\newblock
\APACrefYearMonthDay{2007}{}{}.
\newblock
{\BBOQ}\APACrefatitle {Object categorization: reversals and explanations of the basic-level advantage.} {Object categorization: reversals and explanations of the basic-level advantage.}{\BBCQ}
\newblock
\APACjournalVolNumPages{Journal of Experimental Psychology: General}{136}{3}{451}.
\PrintBackRefs{\CurrentBib}

\bibitem [\protect \citeauthoryear {%
Rogers%
, Patterson%
, Jefferies%
\BCBL {}\ \BBA {} Ralph%
}{%
Rogers%
\ \protect \BOthers {.}}{%
{\protect \APACyear {2015}}%
}]{%
rogers2015disorders}
\APACinsertmetastar {%
rogers2015disorders}%
\begin{APACrefauthors}%
Rogers, T\BPBI T.%
, Patterson, K.%
, Jefferies, E.%
\BCBL {}\ \BBA {} Ralph, M\BPBI A\BPBI L.%
\end{APACrefauthors}%
\unskip\
\newblock
\APACrefYearMonthDay{2015}{}{}.
\newblock
{\BBOQ}\APACrefatitle {Disorders of representation and control in semantic cognition: Effects of familiarity, typicality, and specificity} {Disorders of representation and control in semantic cognition: Effects of familiarity, typicality, and specificity}.{\BBCQ}
\newblock
\APACjournalVolNumPages{Neuropsychologia}{76}{}{220--239}.
\PrintBackRefs{\CurrentBib}

\bibitem [\protect \citeauthoryear {%
Rouse%
\ \protect \BOthers {.}}{%
Rouse%
\ \protect \BOthers {.}}{%
{\protect \APACyear {2023}}%
}]{%
rouse2023neuropsychological}
\APACinsertmetastar {%
rouse2023neuropsychological}%
\begin{APACrefauthors}%
Rouse, M\BPBI A.%
, Halai, A\BPBI D.%
, Ramanan, S.%
, Patterson, K.%
, Rowe, J\BPBI B.%
\BCBL {}\ \BBA {} Ralph, M\BPBI A\BPBI L.%
\end{APACrefauthors}%
\unskip\
\newblock
\APACrefYearMonthDay{2023}{}{}.
\newblock
{\BBOQ}\APACrefatitle {A neuropsychological investigation of social-semantic knowledge in frontotemporal dementia} {A neuropsychological investigation of social-semantic knowledge in frontotemporal dementia}.{\BBCQ}
\newblock
\APACjournalVolNumPages{Alzheimer's \& Dementia}{19}{}{e082444}.
\PrintBackRefs{\CurrentBib}

\bibitem [\protect \citeauthoryear {%
Sievert%
, Nowak%
\BCBL {}\ \BBA {} Rogers%
}{%
Sievert%
\ \protect \BOthers {.}}{%
{\protect \APACyear {2023}}%
}]{%
sievert2023efficiently}
\APACinsertmetastar {%
sievert2023efficiently}%
\begin{APACrefauthors}%
Sievert, S.%
, Nowak, R.%
\BCBL {}\ \BBA {} Rogers, T.%
\end{APACrefauthors}%
\unskip\
\newblock
\APACrefYearMonthDay{2023}{}{}.
\newblock
{\BBOQ}\APACrefatitle {Efficiently Learning Relative Similarity Embeddings with Crowdsourcing} {Efficiently learning relative similarity embeddings with crowdsourcing}.{\BBCQ}
\newblock
\APACjournalVolNumPages{Journal of Open Source Software}{8}{84}{4517}.
\PrintBackRefs{\CurrentBib}

\bibitem [\protect \citeauthoryear {%
Zahn%
\ \protect \BOthers {.}}{%
Zahn%
\ \protect \BOthers {.}}{%
{\protect \APACyear {2007}}%
}]{%
zahn2007social}
\APACinsertmetastar {%
zahn2007social}%
\begin{APACrefauthors}%
Zahn, R.%
, Moll, J.%
, Krueger, F.%
, Huey, E\BPBI D.%
, Garrido, G.%
\BCBL {}\ \BBA {} Grafman, J.%
\end{APACrefauthors}%
\unskip\
\newblock
\APACrefYearMonthDay{2007}{}{}.
\newblock
{\BBOQ}\APACrefatitle {Social concepts are represented in the superior anterior temporal cortex} {Social concepts are represented in the superior anterior temporal cortex}.{\BBCQ}
\newblock
\APACjournalVolNumPages{Proceedings of the National Academy of Sciences}{104}{15}{6430--6435}.
\PrintBackRefs{\CurrentBib}

\end{thebibliography}

\end{document}